\documentclass[letterpaper,12pt,notoc]{JHEP3}

\usepackage{graphicx}
\usepackage{amsmath}
\usepackage{amssymb}

\def\cN{\mathcal{N}}

\def\cO{\mathcal{O}}

\def\tr{\mathop{\mathrm{tr}}}
\def\Im{\mathop{\mathrm{Im}}}

\def\vev#1{\langle#1\rangle}
\def\half{\hbox{$\frac 12$}}

\preprint{
\hbox{}\hfill arXiv:0809.3238}

\title{A counterexample to the {\boldmath$a$}-\textrm{`}theorem\textrm{'}}

\author{
Alfred D. Shapere$^{1,2}$ and Yuji Tachikawa$^2$\\

\bigskip

$^1$ Department of Physics and Astronomy, University of Kentucky, \\
Lexington, Kentucky 40506-0055, USA

\medskip

$^2$ School of Natural Sciences, Institute for Advanced Study,\\
 Princeton,  New Jersey 08540, USA
}

\abstract{
The conclusion of the paper was wrong, due to the incorrect assumption that the low-energy limit at the strongly-coupled point consists of a single, coupled SCFT.  By taking into account the fact that the low-energy limit consists of multiple decoupled parts, it was later shown in \cite{GST} that there is no violation of the $a$-theorem in this system. Furthermore, the $a$-theorem itself was convincingly demonstrated in \cite{KS}, and the argument presented there has been further refined. 
The rest of this paper is kept as it was, for some parts of the discussions might still be of interest.

\bigskip

\noindent\textsc{Original abstract:} We exhibit a renormalization group flow for a four-dimensional gauge theory along which the conformal central charge $a$ increases. The flow connects the maximally superconformal point of an $\cN=2$ gauge theory with gauge group SU$(N{+}1)$ and $N_f=2N$ flavors in the ultraviolet, 
to a strongly-coupled superconformal point of the SU$(N)$ gauge theory with $N_f=2N$ massless flavors in the infrared.
Our example does not contradict the proof of the $a$-theorem via $a$-maximization,
due to the presence of accidental symmetries in the infrared limit.
Nor does it contradict the holographic $a$-theorem, because these gauge theories do not possess weakly-curved holographic duals.
}


\begin{document}

\section{Introduction}
\label{introduction}
Zamolodchikov's  $c$-theorem \cite{Zc} is one of the central results of two-dimensional quantum field theory.  It extends the conformal central charge $c$, defined for conformal field theories, to a function on the space of two-dimensional field theories. This function decreases along renormalization group (RG) flows and is stationary at RG fixed points. 

Over the past two decades, much effort has gone into seeking an analogue of the $c$-theorem in four dimensions.  This effort has been  complicated by the fact that comparatively little is known about nontrivial 4D conformal field theories.  What is known is for the most part limited to superconformal field theories.

In two dimensions, the conformal central charge is proportional to the trace anomaly
in a curved background
\begin{equation}
\vev{T_\mu^\mu} = -{c\over 12}R. \label{twodtrace}
\end{equation}
Similarly, in 4D superconformal field theories the trace anomaly depends on two constants, $a$ and $c$:
\begin{equation}
\vev{T_\mu^\mu} = {c\over 16\pi^2}(\text{Weyl})^2 - {a\over 16 \pi^2}(\text{Euler})\label{traceanomaly}
\end{equation}
where
\begin{eqnarray}
({\rm Weyl})^2 &=& R^2_{\mu\nu\rho\sigma}-2R^2_{\mu\nu}+{1\over 3} R^2,\\
({\rm Euler})&=& R^2_{\mu\nu\rho\sigma}-4R^2_{\mu\nu}+ R^2.
\end{eqnarray}
Like their two-dimensional cousin, the conformal central charges $a$ and $c$ also appear in the stress-tensor OPE.

It is natural to ask whether $a$,  $c$, or some linear combination of them, decreases along all RG flows.   Since $a$ and $c$ are  defined by the above equation only at a CFT, we should ask more specifically whether  $a_{\rm UV} > a_{\rm IR}$ or $c_{\rm UV} > c_{\rm IR}$, where for example $a_{\rm UV}$ denotes the value of $a$ at the UV fixed point of the flow.  (A stronger conjecture
along the lines of the original $c$-theorem,
which posits an interpolating monotonic function,
will not be needed here since we shall find even this weaker conjecture to be false.)
By computing $a$ and $c$ for various pairs of ${\cal N}=1$ superconformal field theories (SCFTs)  connected by RG flows, Anselmi et al. showed that no such statement is true for $c$, nor for any linear combination of $a$ and $c$ other than, possibly, $a$ itself \cite{Anselmi2}.   In all cases they were able to check, they found that $a_{\rm UV} > a_{\rm IR}$.

Further evidence and a general argument in support of an $a$-`theorem' were given by Intriligator and Wecht \cite{IW}, using the $a$-maximization prescription. As they noted, their argument relies on at least two assumptions: first, that no ``accidental'' U(1) symmetries that could potentially mix with U(1)$_R$ appear in the IR, and second, that the local maximum of $a$ in the UV implied by $a$-maximization is actually a global maximum along the flow.  In the works \cite{Kutasov1,Kutasov2,IW2}, the effects of accidental U(1) symmetries which appear when chiral composite operators hit the unitarity bound was taken into account,  but  there are many known examples with $\cN=2$ supersymmetry where other types of accidental U(1) symmetries appear in the IR limit. This left a big loophole in the $a$-theorem.

In this paper, we will give examples of  RG flows that violate the conjectured $a$-theorem without contradicting existing results.  We will employ the method of \cite{ST} for calculating $a$ and $c$ in $\cN=2$ superconformal field theories, known as Argyres-Douglas (AD) points, which are realized as fixed points of $\cN=2$ gauge theories \cite{AD}.  In particular, we will study $\cN=2$ SCFTs of maximal rank, which arise in SU$(N_c)$ gauge theories with $N_f$ fundamental flavors \cite{APSW,EHIY}.  Pairs of these SCFTs are linked by renormalization group flows, along which $a$ should decrease if the $a$-theorem is valid.  Instead, we will find examples of flows for which $a_{\rm UV} < a_{\rm IR}$.  Specifically, these flows connect SCFTs of maximal rank, along which $N_c$ decreases but $N_f$ is unchanged.

The example we will consider in detail is an RG flow from the maximal-rank AD point of SU$(N{+}1)$ gauge theory with $N_f=2N$ quark flavors to the superconformal SU$(N)$ theory  with $2N$ massless flavors at infinitely strong coupling.
We will show that the central charges $a$ at the UV and IR endpoints of this flow are \begin{equation}
a_{UV}=\frac{14N^2+19N}{72}\ ,\qquad
a_{IR}=\frac{7N^2-5}{24}\ , \label{error1}
\end{equation} which violates the $a$-theorem when $N\ge 4$.
In fact, in the large $N$ limit, $a_{UV}\sim(2/3) a_{IR}$, giving a violation of the $a$-theorem of order $N^2$.

We will exhibit a specific deformation of the former theory which flows in the IR to the latter.  The SU$(N)$ theory in the IR possesses a marginal coupling $\tau$;  when the deformation of the SU$(N{+}1)$ theory is small, the IR endpoint of the flow is at very strong coupling $\tau \sim 1$.

Another notable property of this flow is the behavior of the dimensions of the Coulomb branch operators.   In the UV are they given by \begin{equation}
D_{UV}(u_{j})=\frac{2}{3} j,
\end{equation} while the corresponding dimensions in the IR SCFT are \begin{equation}
D_{IR}(u_{j})=j-1.
\end{equation} This means that in general the dimension of each operator in the IR is significantly larger than the dimension of the corresponding operator in the UV, which is contrary to the behavior of perturbative Banks-Zaks type theories \cite{BZ}, where the gauge interaction plays a dominant role in creating the superconformal point.  It is reasonable to attribute at least some part of the increase in $a$ to this
large increase in anomalous dimensions.  An indication that this is indeed the case comes from the general formula  \cite{AW,ST}
\begin{equation}
4(2a-c)=\sum_{i=1}^r (2D(u_i)-1)
\end{equation} valid in all  4d $\cN=2$ SCFTs under consideration. Inserting the above anomalous dimensions into the sum on the right-hand side of this equation, it is easy to see that the combination $2a-c$, like $a$, increases by a factor of approximately 3/2 along the flow in question. 

There are many other examples of flows of this type.
From the formula for $a$ that we will derive in Sec.~\ref{calculation},
it follows directly that for any $N_c$ and $N_f$ satisfying \begin{equation}
2N_c > N_f > \hbox{$\frac12$}(\sqrt{21}-3)N_c\approx 0.79 N_c \ , \end{equation}
with $N_c$ sufficiently large, there is a flow between Argyres-Douglas points of the SU($N_c{+}1$) and SU($N_c$) gauge theories, both with $N_f$ flavors, along which $a$ increases.

The first question raised by our result is, how do our examples avoid
falling into one of the classes of theories for
which the $a$-theorem has been established.
First, as has already been mentioned, the  the proof using the method of $a$-maximization  \cite{Kutasov1,Kutasov2,IW2} relies on a strong assumption about the non-existence of accidental symmetries.  In our case, this assumption is clearly
violated: the
U$(1)_R$ symmetry is broken throughout the RG flow, and the
U$(1)_R$ that appears in the IR has nothing to do with the UV $R$-symmetry;  it 
is totally accidental.  
Second, our examples do not contradict the holographic derivation of the $a$-theorem because the CFTs involved have no holographic duals. Indeed, any CFT with a weakly curved gravity dual
has $a$ and $c$ both of order $N^2$ and $a-c$ of order at most $N$ \cite{HenningsonSkenderis}, whereas  in our case $a-c$ is of order $N^2$.
Third, it might seem rather surprising  to the reader that counterexamples can be found
within a class of SCFTs that was first discussed in 1996 \cite{EHIY},
well before many of the more modern analyses of the $a$-theorem  \cite{IW,holographic-c}.
The reason these counterexamples were not recognized sooner is simply that there was no method to calculate $a$ for these theories before the work \cite{ST}.

We begin  by reviewing in Sec.~\ref{review} the method developed in \cite{ST} for obtaining the central charges $a$ and $c$ of $\cN=2$ SCFTs .
We apply this method in Sec.~\ref{calculation} to the maximal-rank superconformal points of SU$(N_c)$ gauge theory with $N_f$ flavors, which were first studied in \cite{EHIY}.
In Sec.~\ref{violation} we will find
that there is a violation of the $a$-theorem within this class of SCFTs, and
we will take some care to establish that there is indeed an RG flow between the two particular SCFTs involved in our main example.  We close the paper  in Sec.~\ref{discussion} with a discussion of our results.
 We briefly summarize the history of the four-dimensional analogue of the $c$-theorem in Appendix~\ref{history}.

\section{$a$ and $c$ for $\cN=2$ gauge theories}
\label{review}
Recently we developed a method for calculating central charges of $\cN=2$ gauge theories, by relating them to $R$-symmetry anomalies in the corresponding topological field theory \cite{ST}.  Here we will review the method, in order to set the stage for its application in the next section to the calculation of $a$ and $c$ for SU($N$) gauge theories with hypermultiplet matter.

We begin by recalling the relation of $a$ and $c$ to the anomalous conservation law of the
U(1)$_R$ current in $R^\mu$ any $\cN=2$ field theory \cite{KT},  in the presence of a background metric and a background SU(2)$_R$ gauge field $F_{\mu\nu}^a$:
 \begin{equation}
\partial_\mu R^\mu=
\frac{c-a}{8\pi^2}
R_{\mu\nu\rho\sigma}\tilde R^{\mu\nu\rho\sigma}
+
\frac{2a-c}{8\pi^2}F_{\mu\nu}^{a}\tilde F^{\mu\nu}_{a}\label{cN=2-relations}
\end{equation}  
where \begin{equation}
\tilde F^a_{\mu\nu}=\half\epsilon_{\mu\nu\rho\sigma}F^a_{\rho\sigma},\qquad
\tilde R_{\mu\nu\rho\sigma}=\half\epsilon_{\mu\nu \alpha\beta}R_{\alpha\beta\rho\sigma}.
\end{equation}
According to the well-known construction of topological gauge theories, 
in backgrounds where the SU(2)$_R$ gauge field is equal to the self-dual part of the curvature, {\it i.e.}
\begin{equation}
F^a_{\mu\nu}t^a_{\rho\sigma}=
\half (R_{\mu\nu\rho\sigma}+\tilde R_{\mu\nu\rho\sigma}) \label{twist}
\end{equation} with the 't Hooft symbol $t^a_{\rho\sigma}$,
correlation functions of physical operators depend only on the topology of the background manifold.
Substituting this condition into \eqref{cN=2-relations} gives the anomaly equation in the topological background, which when integrated over the 4-manifold gives
 the total $R$-charge of the vacuum
\begin{equation}
\Delta R =  2(2a-c) \chi+ 3c\,\sigma \label{a-c-anomaly}
\end{equation} in terms of the Euler characteristic $\chi$ and the signature $\sigma$ of the manifold.

Thus to determine $a$ and $c$, it suffices to be able to compute the dependence of $\Delta R$ on the topology of the background.  This information is encoded in the path integral measure, which for a topological gauge theory takes the form \begin{equation}
[d\mu] A^\chi B^\sigma \label{measurefactors}
\end{equation}
The factor $[d\mu]$ is the measure for the $r$ vector multiplets, which at a generic point in moduli space are the only massless modes.  
The measure factors $A$ and $B$ depend holomorphically on the Coulomb branch moduli and are associated with the additional massless states that appear on special loci of complex codimension 1 and higher. 

The $R$-charge of the vacuum can be directly read off from the measure \eqref{measurefactors}
\begin{equation}
\Delta R = \chi R(A) + \sigma R(B) + \frac{\chi+\sigma}{2} r .
\label{DeltaR}
\end{equation}
Here $R(A)$ and $R(B)$ denote the $R$ charges of $A$ and $B$, and the last term is the contribution of the generically massless vector multiplets.
Comparing with \eqref{a-c-anomaly}, we find the following general expressions for $a$ and $c$:
\begin{align}
a&=\frac14R(A)+\frac16R(B)+\frac5{24}r,&
c&=\frac13R(B)+\frac1{6}r\label{master-formula}
\end{align}

Finding $a$ and $c$ is thus reduced to calculating the $R$-charge of the functions $A$ and $B$.  In general, these functions are believed to take the form
\cite{W,MW,LNS1,MM,MMP}
\begin{equation}
A(u)=\alpha\left[\det \frac{\partial u_i}{\partial a^I}\right]^{1/2},\qquad
B(u)=\beta\Delta^{1/8}.\label{generalconjecture}
\end{equation}
Here, $u_i$ are gauge- and monodromy- invariant coordinates
on the Coulomb branch, $a^I$ are special coordinates,
and $\Delta$  is the physical discriminant of the Seiberg-Witten curve.
$\alpha$ and $\beta$ are prefactors independent of the $u_i$ which
can in principle depend on the mass parameters.  The functions in \eqref{generalconjecture}
are readily computable in the vicinity of
many superconformal points of $\cN=2$ gauge theories.

The $R$-charges of $A$ and $B$ can be written in terms of the $R$-charges $R(u_i)$,
or their dimensions $D(u_i)$, which satisfy  \begin{equation}
R(u_i)=2D(u_i)
\end{equation} by virtue of superconformal symmetry. Finally, the dimensions $D(u_i)$ can be obtained from the scaling form of the Seiberg-Witten curve by demanding that the dimension of the Seiberg-Witten
differential $\lambda_{SW}$ be one. This completes the calculation of the central charges $a$ and $c$.

\section{Superconformal points of $\cN=2$  SU($N_c$) with quarks}
\label{calculation}
We will now apply the method described in the previous section to calculate the conformal central charges of an infinite family of 4D SCFTs, which 
correspond to Argyres-Douglas points \cite{AD}  in the Coulomb branch of $\cN=2$ SU$(N_c)$ gauge theory with $N_f$ flavors of fundamental quarks \cite{EHIY}.  At these points, a maximal set of mutually nonlocal dyons becomes massless, and the Seiberg-Witten curve develops a singularity of maximal rank.
Here the rank of a superconformal theory signifies the minimal 
number of U(1) vector multiplets
to which the set of dyons with degenerating mass couple electrically or magnetically.
We will restrict our discussion to theories with even $N_f\equiv 2n_f$.

To locate these points, we start with the Seiberg-Witten curve for $\cN=2$ supersymmetric SU($N_c$) gauge theory with $N_f< 2N_c$ fundamental hypermultiplets of equal mass, which has the form \cite{APS,HO}
\begin{equation}
y^2=P(x)^2 - \Lambda^{2N_c-N_f} (x+m)^{N_f} \label{curve}
\end{equation}
where \begin{equation}
P(x) = x^{N_c}+u_{2}x^{N_c-2} + u_{3} x^{N_c-3}\cdots +u_{N_c}.
\end{equation} $\Lambda$ is the dynamically generated scale of the gauge theory,
and  one can identify $u_j$ with the composite operator $\tr \phi^j$
in the semiclassical regime.
The Seiberg-Witten differential is \begin{equation}
\lambda_{SW}=x d\log \frac{1-y/P}{1+y/P}. \label{oneform}
\end{equation}

To reach the superconformal point of maximal rank, we first choose the moduli $u_{i}$
so that  \begin{equation}
P(x)=(x+m)^{n_f} C_{N_c-n_f}(x)
\end{equation}
where $C_{N_c-n_f}(x)$ is a polynomial of degree $N_c-n_f$.
Then the curve becomes
\begin{equation}
y^2=(x+m)^{N_f}( C_{N_c-n_f}(x) - \Lambda^{N_c-n_f})(C_{N_c-n_f}(x) + \Lambda^{N_c-n_f})
\end{equation}
The roots of $C_{N_c-n_f}$ can further be adjusted by tuning the remaining moduli and $m$; we can use this freedom to set
$C_{N_c-n_f}(x)  = (x+m)^{N_c-n_f} - \Lambda^{N_c-n_f}$, giving a singularity of maximal degree
\begin{equation}
y^2=(x+m)^{N_c+n_f}((x+m)^{N_c-n_f} - 2\Lambda^{N_c-n_f}).
\end{equation}

For $N_c-n_f\ge 2$  this procedure leads to the choice \begin{equation}
m=0,\qquad P(x)=x^{N_c}-\Lambda^{N_c-n_f} x^{n_f}.
\end{equation}
It is known that another branch of the moduli space touches the Coulomb branch at this point.  In the terminology of \cite{higgs}, this is a special point on the non-baryonic Higgs branch root with extra massless monopoles, whose generic massless spectrum is that of a U$(n_f)$ gauge theory with $N_f=2n_f$ quarks.
The Higgs branch emanating from this point has quaternionic dimension $n_f^2$.

For $N_c-n_f=1$ we need to choose \begin{equation}
m=\frac{\Lambda}{N_c},\qquad P(x)=(x+m)^{N_c}-\Lambda x^{N_c-1}
\end{equation} in order to guarantee that the coefficient of the $x^{N_c-1}$ term of $P(x)$ vanishes.
Again this point lies at the root of a non-baryonic Higgs branch
of quaternionic dimension $n_f^2$.

The SCFT at this point was first studied by \cite{EHIY} and denoted by the symbol $M^{N_f}_{N_c+n_f}$.
By expanding the one-form \eqref{oneform} around this point and demanding $D(\lambda_{SW})=1$,
the authors of \cite{EHIY} found the scaling dimensions \begin{equation}
D(x)=\frac{2}{N_c-n_f+2},\qquad
D(u_{j})= j D(x).\label{dimensions}
\end{equation}

When $N_c+n_f$ is odd,
there are $r=(N_c+n_f-1)/2$
pairs of special coordinates $a^I$  which become zero at the superconformal point,
{\it i.e.}~the rank of this theory is $r$.
The Coulomb branch operators with dimension $>1$
are $u_j$ with $r-n_f+2\le j \le N_c$.
So there are  $N_c-(r-n_f+2)+1=r$ of them, as expected.
As was argued in \cite{EHIY}, there are loci in the SU$(N_c)$ gauge theory with $N_f$ quarks
where the low-energy theory becomes superconformal with non-maximal rank $r'<r$.
These non-maximal superconformal points with rank $r'$ are known to be equivalent
to the maximal-rank superconformal points of the SU$(N_c-2r+2r')$  gauge theory with $N_f$ quarks. Thus the maximal superconformal points are naturally labeled by their rank $r$ and the number of flavors $N_f$, and we will express the central charges in terms of these quantities.

The dimensions \eqref{dimensions} determine the $R$-charge of the measure factor $A$:
\begin{equation}
R(A)=\sum_{j=r-n_f+2}^{N_c} \left(D(u_{j})-1\right)=\frac{r^2}{2r-2n_f+3}.
\end{equation}
Also, the discriminant of the curve is
\begin{equation}
\Delta=B^8=\prod_{i>j}(e_i-e_j)^2,
\end{equation} where $e_i$ ($i=1,\ldots,2N$) are the branch points of the curve \eqref{curve}.
Note that only $2r+1$ of the branch points $e_i$ are small and
have the same dimension as $\delta x$.
Thus we have
\begin{equation}
R(B)=\frac{r(2r+1)}{2r-2n_f+3}.
\end{equation} Therefore the central charges are given by the formula \eqref{master-formula} : \begin{equation}
a=\frac{r(24r-10n_f+19)}{24(2r-2n_f+3)},\quad
c=\frac{r(6r-2n_f+5)}{6(2r-2n_f+3)}.
\end{equation}

Similarly, when $N_c+n_f$ is even,
there are $r=(N_c+n_f)/2-1$ pairs of special coordinates $a^I$  which become zero.
The Coulomb branch operators with dimension $>1$
are $u_j$ with $r-n_f+3\le j \le N_c$.
So there are again $N_c-(r-n_f+2)+1=r$ of them, as expected.
We have \begin{equation}
R(A)=\frac{r(r+1)}{2(r-n_f+2)},\quad
R(B)=\frac{(r+1)(2r+1)}{2(r-n_f+2)}.
\end{equation} Thus the central charges are \begin{equation}
a=\frac{12r^2+(19-5n_f)r+2}{24(r-n_f+2)},\quad
c=\frac{3r^2+(5-n_f)r+1}{6(r-n_f+2)}.
\end{equation} Note that in all cases, the ratio of the central charges satisfies the inequality \begin{equation}
\frac12\le \frac{a}{c}\le \frac54
\end{equation} which was discussed in \cite{DiegoJuan,ST}.

Let us study the case $N_f=2N_c$  separately. The curve is given by \cite{APS}
\begin{equation}
y^2 = P(x)^2- f(\tau) Q(x),\label{Nf=2N}
\end{equation}where \begin{align}
P(x)& = x^{N_c}+u_{2}x^{N_c-2} + u_{3} x^{N_c-3}\cdots +u_{N_c},
\label{Px}\\
Q(x)&=\prod_{a=1}^{2N_c} (x-m - 2g(\tau)m).
\end{align} Here,
 $f(\tau)$ and $g(\tau)$ are certain modular functions of the complexified gauge coupling $\tau$.
The superconformal point of maximal rank $r=N_c-1$ is reached by scaling the mass $m$ and vevs $u_{k}$ to zero with canonical scaling dimensions.
This conformal point has one marginal coupling $\tau$.
The central charges we obtain from the formula \eqref{master-formula} are
\begin{equation}
a = \frac7{24}N_c^2-\frac5{24},\qquad
c =\frac13 N_c^2-\frac16.\label{free-ac}
\end{equation} Alternatively, the central charges can be computed in the weak-coupling limit $\Im \tau\to \infty$ as sums of contributions of free fields.  Since $a$ and $c$ must be independent of $\tau$, the free-field answer should, and does, agree with \eqref{free-ac}.

\section{Violation of the $a$-`theorem'}
\label{violation}

Let us now compare the central charges $a$ of two of the SCFTs studied above, 
the SU$(N+1)$ theory
and the SU$(N)$ theory, both with $N_f=2N$ quark flavors.
We claim that there is a renormalization group flow from the former to the latter.
The values of the central charge $a$ at the UV and IR endpoints of this flow are then\begin{equation}
a_{UV}=\frac{14N^2+19N}{72}\ ,\qquad
a_{IR}=\frac{7N^2-5}{24}\ , \label{error2}
\end{equation} which violates the $a$-theorem when $N\ge 4$.
In fact, in the large $N$ limit $a_{UV}\sim (7/36) N^2$ and $a_{IR}\sim (7/24)N^2$
so we have $a_{UV}\sim(2/3) a_{IR}$, which amounts
to an  $a$-theorem violation of $\cO(N^2)$.

In order to confirm that we have indeed found a counterexample to the $a$-theorem, we need to establish that there is in fact an RG flow starting from the
maximal superconformal point of the SU$(N{+}1)$ theory with $2N$ quarks in the ultraviolet, to the SU$(N)$ theory with the same number of quarks.

To this end, let us study the SU$(N{+}1)$ theory with $2N$ quarks in more detail.
When $P(x)=(x+m)^{N}(x-Nm)$, the curve becomes \begin{equation}
y^2=(x+m)^{2N}(x-Nm+\Lambda)(x-Nm-\Lambda).
\end{equation} Therefore the maximal-rank superconformal point
occurs when $-Nm+\Lambda=m$,
as also discussed in the previous section.
We parameterize  the deviation from this value of $m$ as \begin{equation}
-Nm+\Lambda= m + \delta m
\end{equation} and expand $P(x)$ around the superconformal point as \begin{align}
P(x)&=(x+m)^{N}(x-Nm)+ \tilde u_2 (x+m)^{N-1}+ \cdots\\
&=\tilde x^N(\tilde x +\delta m -\Lambda )+\tilde u_2 \tilde x^{N-1} +\tilde u_3\tilde x^{N-2}+\cdots.
\end{align} where $\tilde x=x+m$.
Now the curve takes the form \begin{multline}
y^2=\left[x^N(x+\delta m)+u_2 x^{N-1}+u_3x^{N-2}+\cdots\right]\\
\times\left[x^N(x+\delta m-2\Lambda)+u_2 x^{N-1}+u_3x^{N-2}+\cdots\right]\label{ADcurve}
\end{multline} where we have dropped all tildes. Note that we have not yet made any approximations.

Let us now study the behavior of the curve \eqref{ADcurve} very close to the superconformal point, in the limit \begin{equation}
|x|\sim |u_j|^{1/j}\sim |\delta m|  \ll |\Lambda|
\end{equation} Then the curve is approximately \begin{equation}
y^2 \sim -2\Lambda x^N \left[ x^N(x+\delta m) +u_2 x^{N-1}+u_3x^{N-2}+\cdots\right].
\end{equation} This describes a general deformation of the maximal superconformal point. 

If we instead take the scaling limit \begin{equation}
|x|\sim |u_j|^{1/(j-1)} \ll |\delta m| \ll  |\Lambda| ,
\end{equation} then the curve becomes approximately \begin{align}
y^2&\sim \left[x^N(\delta m)+u_2 x^{N-1}+u_3x^{N-2}+\cdots\right]\nonumber\\
&\qquad\times\left[x^N(\delta m-2\Lambda)+u_2 x^{N-1}+u_3x^{N-2}+\cdots\right]\\
&= \left[ (\delta m -\Lambda) x^N+u_2 x^{N-1} +u_3 x^{N-2}+\cdots \right] ^2 - \Lambda^2 x^{2N}
\end{align} 
Absorbing the factor $(\delta m - \Lambda)$ into $y$, shifting $x$ to eliminate the second term in brackets, and redefining $\hat u_j \equiv u_{j+1}/(\delta m - \Lambda)$, we finally obtain\begin{align}
y^2&\sim \left[ x^N+{\hat u}_2 x^{N-2}+\cdots \right] ^2 - \left(\frac{\Lambda}{\delta m - \Lambda}\right)^2 (x-\mu)^{2N}
\end{align} 
which we recognize as the curve \eqref{Nf=2N} of the SU$(N)$ theory with $2N$  flavors of mass 
\begin{equation}
\mu\equiv\frac{u_2}{N(\delta m - \Lambda)}\end{equation} 
at a particular value of the coupling $\tau$ 
depending on the ratio of $(\delta m-\Lambda)$ and $\Lambda$. 
The mass is automatically zero when $u_2=0$, and then
the change in $\delta m$ directly translates to a change in $\tau$.
 Using the explicit form of the modular function $f(\tau)$ \cite{APS}, this value is found to be close to the infinite coupling point $\tau=1$. 

The subspace of the SU$(N)$ moduli-parameter space generated by deforming the $\tau=1$ superconformal point by $\delta m$ extends out to the semiclassical region. There, it can be matched onto the moduli space of the parent SU$(N{+}1)$ theory with $2N$ quarks of mass $m$, Higgsed down to the SU$(N)$ theory with $2N$ flavors by the adjoint scalar vev \begin{equation}
\vev{\phi} = \text{diag}\ (m,m,\ldots,m,-Nm).
\end{equation}

\FIGURE{
\centerline{\includegraphics[width=.6\textwidth]{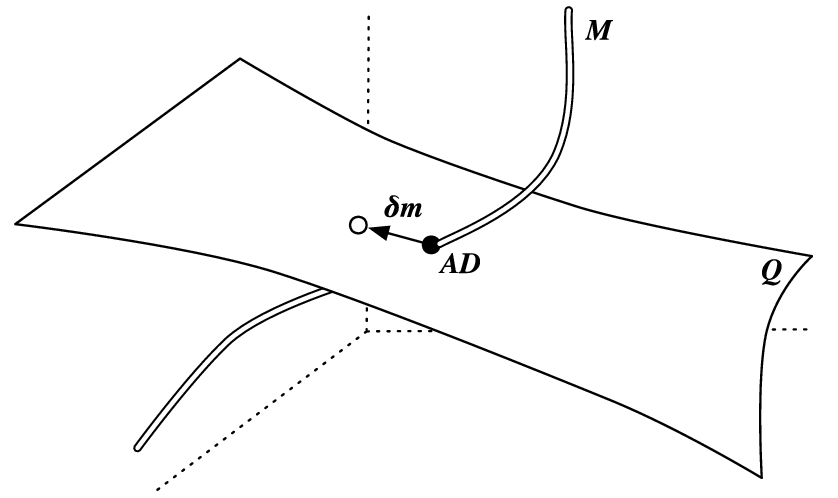}}
\caption{Depiction of the moduli-parameter space of  the SU$(N{+}1)$ theory
with $2N$ flavors. $Q$ is the subspace where the low energy theory
has unbroken SU$(N)$ gauge bosons and
$2N$ massless flavors. 
The locus $M$ where an extra monopole becomes massless intersects with $Q$ at the AD point.
\label{moduli}}
}

A schematic picture of the moduli-parameter space of the SU$(N{+}1)$ gauge theory is depicted in Fig.~\ref{moduli}.
There, the subspace $Q$ is the locus generated by $\delta m$ inside the space of $m$, $u_2$, \ldots, $u_{N+1}$
where the low energy theory contains massless SU$(N)$ gauge bosons and $2N$ massless flavors.
(It is slightly unconventional to depict the mass parameter $m$ and the moduli $u_j$ together in the same moduli-parameter space, 
but as was explained in \cite{APSW},
this is a natural point of view to take in our situation.)
Indeed, one may think of the space of $m$ and $u_j$ as the moduli space of a
U$(N{+}1)$ gauge theory with $2N$ flavors.
The locus $M$, where an extra magnetically charged state becomes massless,
intersects with $Q$ at the AD point.
This depiction is highly schematic, {\it e.g.}~in that the intersection is not transversal as in the figure
and is far more complicated in reality, as
was well-illustrated for the case $N=2$ in the original paper \cite{EHIY}.
(See the discussion following Eq.(33) therein.)

With these preparations, fix $|\delta m| \ll |\Lambda|$, and consider the following three regimes of vevs of $u_j$'s:
\begin{align}
{\mathbf 1}.\ & |u_j|\sim |\Lambda|^j;&
{\mathbf 2}.\ & |u_j| \sim |\delta m| ^j;&
{\mathbf 3}.\  & |u_j| \sim \epsilon^{j-1}|\delta m|,\ \epsilon \ll |\delta m| \nonumber
\end{align}
For $u_j$ in regime 1, we are at a generic point in the moduli space of  the SU$(N{+}1)$ theory with $2N$ flavors with no particular interest. The special coordinates $a^I$, or equivalently the masses of the BPS solitons,  are all of order $\Lambda$.
As we lower the $u_j$ and enter regime 2, the system exhibits
the scaling of the maximal AD point of SU$(N{+}1)$ theory with $2N$ flavors, {\it i.e.}~the dimension
of $u_j$ is $2j/3$.
The low-lying spectrum of BPS masses is that of the maximal AD point,
of order $(\delta m)^{3/2}$.
When we further lower $u_j$ to regime 3,
the scaling dimensions
are those of the SU$(N)$ theory with $2N$ massless flavors, and $u_j$ has canonical dimension $j-1$.  The low-lying  BPS solitons have masses of order $\epsilon$.

Our discussion up to this point has just been a standard analysis of a trajectory through the moduli space of an $\cN=2$ gauge theory. In other words, we have studied how the couplings and BPS masses behave under a particular change of the vevs.  While this trajectory is reminiscent of an RG flow, in that the energy scale defined by the vevs is changing, the true RG flow is along an extra direction which is not tangent to the moduli space. Indeed, the solution of $\cN=2$ gauge theories via Seiberg-Witten curves describes the moduli-dependence of infrared fixed points, which by definition do not flow. 

\FIGURE{
\centerline{
\begin{tabular}{c@{\qquad\qquad}c}
\includegraphics[width=.3\textwidth]{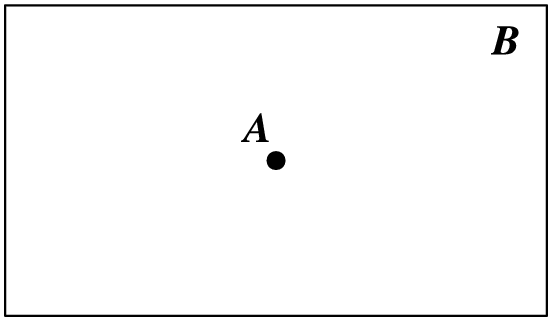} &
\includegraphics[width=.5\textwidth]{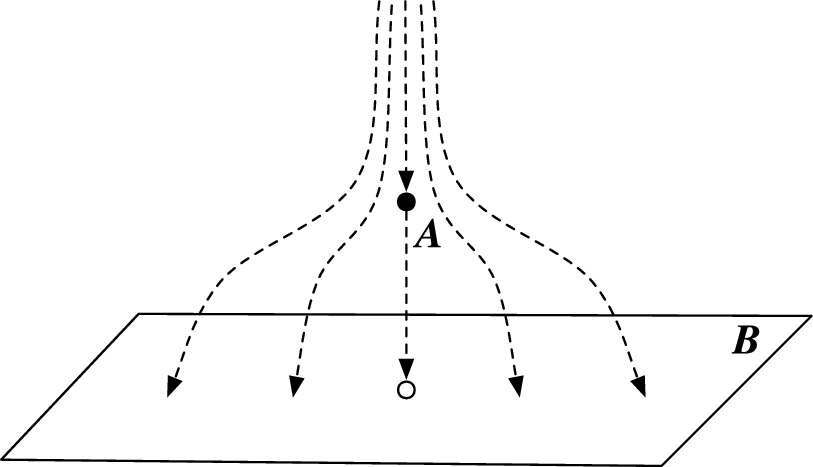} \\
Moduli space & RG flow
\end{tabular}
}
\caption{Distinction between the moduli space and the RG flow.  See the text for explanation. 
\label{distinction}}
}

Let us first understand how RG flow is manifested in the simpler case of the pure SU$(2)$ gauge theory.
The moduli space is parameterized by a single vev $u$, and the low energy theory at generic values of $u$ is a free U$(1)$ gauge theory with coupling $\tau(u)$. The moduli space thus consists of a family of trivial conformal fixed
points parameterized by the marginal coupling $\tau$. 
For special values of $u$, say $u=\Lambda^2$, 
the low-energy fixed point theory includes an extra massless hypermultiplet and the gauge coupling vanishes.
This fixed point can be deformed by the operator $u$ which makes the hypermultiplet massive;
the endpoint of the flow from the deformed theory is a free U$(1)$ gauge theory. 

This situation is heuristically depicted in Fig.~\ref{distinction}.
On the left-hand side, the moduli space is split into a special point $A$, where the spectrum includes an extra massless
hypermultiplet, and the generic region $B$, where the low-energy limit
contains only a U$(1)$ gauge field. The 
right-hand side is a cartoon of the RG flow:  the point $A$ and 
the family $B$ are low-energy endpoints of flows, and are by definition conformal.
They are embedded in a larger space of non-conformal theories,
through which the RG transformation generates flows.
The special point $A$ can be viewed either as an IR fixed point of a flow from the microscopic theory, or as a UV fixed point whose relevant deformations generate flows to IR fixed point theories in $B$.  
At the conformal point $A$, the coupling of the U$(1)$ gauge field is strictly zero.
Therefore, the flow starting exactly at $A$ ends at the zero coupling limit 
of the family $B$.  If the RG flow starts slightly away from $A$, i.e.~if the gauge coupling
of the theory is not strictly zero, then the endpoint of the flow after the decoupling
of the hypermultiplet has nonzero coupling constant which  thus corresponds 
to a generic point of the family $B$.

We will argue that the RG flow between the AD point and the space $Q$ is quite analogous to the example just discussed, with  $A$ representing the AD point and $B$ corresponding to the moduli space $Q$.
Let us fix $u_j^{(0)}$ and $\delta m^{(0)}$ to be sufficiently small, but finite,
compared to the dynamical scale of the gauge theory $|\Lambda|$. 
We then consider the parameterized locus of deformations away from the AD point \begin{equation}
\delta m = \lambda\,\delta m^{(0)},\qquad u_j = \lambda^j  u_j ^{(0)}. 
\label{deformation}
\end{equation} 
We wish to study the RG flow which passes through a point on this locus. At such a point, the lightest massive BPS states set a mass scale $M(\lambda)$, which we can assume is much less than the dynamical scale $|\Lambda|$ of the gauge theory. 
Since the scaling dimension of $\delta m$ close to the AD point is $2/3$, it follows that $M(\lambda)\sim\lambda^{3/2}$ as $\lambda \to 0$.  
Now consider an RG scale $\Lambda_{RG}$ in the range
\begin{equation}
M(\lambda)\ll \Lambda_{RG} \ll |\Lambda| ,
\end{equation} 
At such scales the BPS states, which become exactly massless at the AD point, are effectively massless, and the theory is effectively equivalent to the superconformal AD theory.

Let us next consider the following trajectory in the moduli space
\begin{equation}
\delta m = \delta m^{(0)},\qquad u_j = \epsilon^{j-1}  u_j ^{(0)},
\end{equation} where we take  $\epsilon$ to be very small.
Then the BPS states have two typical mass scales, $M$ determined by $\delta m$
and $\mu(\epsilon)$ determined by $u_j$'s. By the analysis of the Seiberg-Witten curve,
we know $\mu(\epsilon)$ scales as $\mu(\epsilon)\sim \epsilon$ when $\epsilon\to 0$. 
Therefore the system viewed
at the RG scale $\Lambda_{RG}$ in the range \begin{equation}
\mu(\epsilon) \ll \Lambda_{RG} \ll M
\end{equation} is effectively equivalent to the superconformal SU$(N)$ theory with $2N$ massless
flavors.  Below $\mu(\epsilon)$, the vevs $u_j$ break the SU$(N)$ theory down to 
a theory with decoupled U$(1)$ vector multiplets.
Our interpretation of the RG flow is summarized in Fig.~\ref{diagram}.

\FIGURE{
\centerline{\includegraphics[width=.6\textwidth]{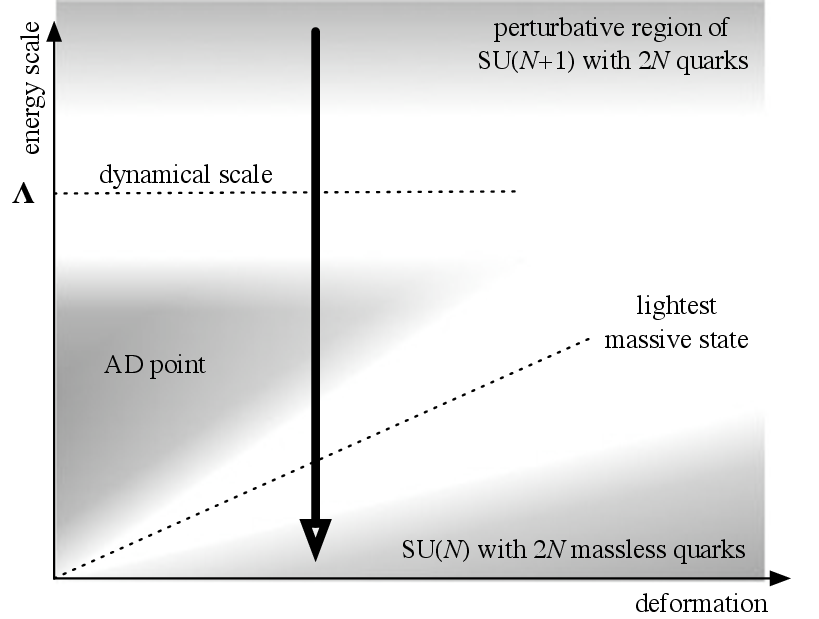}}
\caption{Schematic RG flow.
For scales above $\Lambda$, the system is in the perturbative 
regime of the parent SU$(N{+}1)$ theory with $2N$ flavors.
Below $\Lambda$, the RG flow passes close to the maximal superconformal point.
The deformation  $\delta m$ becomes important once the energy scale is comparable
to the masses of light BPS states, and the theory flows to 
the SU$(N)$ theory with $2N$ massless flavors in the far IR.
The system is nearly conformal in the three shaded regions. 
 \label{diagram}}
}

To recapitulate our discussion, 
the SU$(N{+}1)$ theory with $2N$ quarks and
with $\delta m\ne 0$ 
has the following evolution along the RG flow:
In the extreme UV it  starts as a perturbative gauge theory with gauge group SU$(N{+}1)$
and $2N$ quarks. 
It becomes strongly coupled at a scale of order $\Lambda$,
and gets attracted to the maximal AD point. 
It then starts to be affected by the small deformation $\delta m\ne 0$. 
This deformation is relevant, because the parameter $\delta m$ has dimension $2/3$ and the
corresponding operator which $\delta m$ multiplies in the Lagrangian has dimension  $4/3$.
Far below that scale,
the flow eventually ends at the SU$(N)$ theory with $2N$ massless flavors,
close to the infinite coupling point  $\tau=1$.
Combined with the calculation of $a$ we performed in the last section,
this RG flow establishes the violation of the $a$-theorem.
The behavior of $a$ is shown in Fig.~\ref{graph-of-a}.
This figure is again highly schematic because we do not have a proper interpolating $a$-function at intermediate scales.

\FIGURE{
\centerline{\includegraphics[width=.7\textwidth]{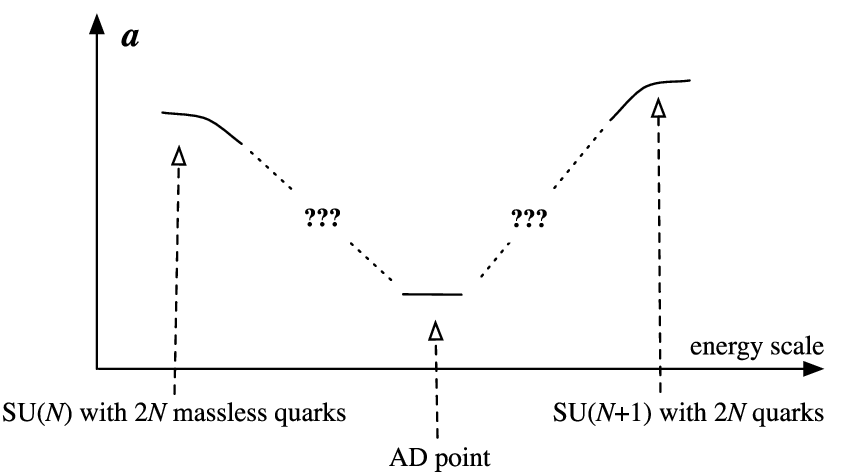}}
\caption{Schematic graph showing the change in the central charge $a$ .\label{graph-of-a}}
}

\section{Discussion}
\label{discussion}

The careful analysis of the preceding section establishes
that there exists an RG flow from the AD point
of the SU$(N{+}1)$  theory with $2N$ quarks to the SU$(N)$ theory with $2N$ massless quarks. If one still wishes to rescue the $a$-theorem, one needs to scrutinize the 
calculation of the central charges in Sec.~\ref{calculation},
which is an application of the authors' recent work \cite{ST}.

One question that could be asked concerns the role of the Higgs branch. In our method, the SCFT point is studied by slightly moving away from it
along the Coulomb branch. But we know that a non-baryonic Higgs branch with quaternionic dimension
$n_f^2$ emanates from the SCFT point, as was recalled in Sec.~\ref{calculation}.
Is the contribution from this branch correctly accounted by this method?
The answer is yes; the factors $A(u)$ and $B(u)$ on the Coulomb branch
arise from integrating out the massive states in the theory. The fact that $A(u)$ and $B(u)$ have zeroes of order $N^2$ at the SCFT point signifies
that the number of light degrees of freedom is of order $N^2$.
In \cite{ST} this method was applied to the USp$(2N)$ theory with $N_f=1,2,3$ quarks, in addition to a hypermultiplet in the antisymmetric representation, which also has a large Higgs branch. The theory has an F-theoretic holographic dual and the Higgs branch corresponds to the absorption of D3-branes onto a stack of 7-branes as instantons.
The fact that the calculation based on this method completely reproduces
the central charges found  using holography in \cite{AT} demonstrates its overall consistency.  There is no problem regarding the possible dependence of the measure factors $A$, $B$ on the Higgs branch
vevs either, because we always work slightly away from the SCFT point along the Coulomb branch, where there are no massless hypermultiplets.

Another question, also related to the Higgs branch, concerns the identification of the U$(1)_R$ symmetry. 
The superconformal U$(1)_R$ charges of the vector multiplet scalars $u_j$
are fixed by the Seiberg-Witten differential,
but there could in principle be extra non-R U$(1)$ symmetries
under which the $u_j$ are neutral, which could mix into the U$(1)_R$. 
An example of such a U$(1)$ symmetry is the U$(1)_B$ symmetry which acts on the Higgs branch 
as the U$(1)$ part of the U$(N_f)$ flavor rotation.  However,
this U$(1)_B$ is vector-like, and thus cannot mix with the U$(1)_R$ symmetry \cite{IW}.  
This conclusion can also be drawn from the relation \eqref{cN=2-relations}:
the central charges are encoded in the 't Hooft anomalies of the forms
U$(1)$-gravity-gravity and U$(1)$-SU$(2)_R$-SU$(2)_R\,$,
but U$(1)_B$ has neither type of anomaly. Thus, it cannot contribute to $a$ or $c$
even if it mixes with the U$(1)_R$ symmetry which we identified.

As was discussed in \cite{higgs}, there is no other U$(1)$ symmetry
which acts on the Higgs branch. The next
possibility to be ruled out is the existence of 
an accidental, non-R, chiral U$(1)$ symmetry (let us call it $T$) 
which appears at the SCFT point, under which the $u_j$ are neutral.
But we find the existence of such a symmetry highly unlikely:
a small, generic deformation along the Coulomb branch
from the SCFT point, which is generated by giving vevs to the $u_j$, does not break $T$ because the $u_j$ are neutral under $T$. 
In other words, this unbroken  symmetry $T$ should also be present slightly away from the SCFT point, where it can act only on massive states because the only massless states away from the SCFT point are free vector multiplets. 
This is a contradiction, because a chiral symmetry $T$ can only act on massless states.\footnote{
Let us apply this argument to a trivial SCFT point whose low energy content
is a U$(1)$ gauge theory coupled to a hypermultiplet, formed by two $\cN=1$ chiral superfields
$(q,\tilde q)$. Denote the vector multiplet scalar by $\phi$.
The theory is completely free in the IR,
so there is an accidental chiral U$(1)$ symmetry, call it $T'$, under which 
both $q$ and $\tilde q$ have charge $+1$, and $\phi$ is neutral.
This $T'$ is broken along the Coulomb branch, which at first appears to contradict the argument presented above.

The point is that this U$(1)$ symmetry does not commute with the SU$(2)_R$ symmetry.
Existence of such symmetries is usually forbidden by the Haag-\L opusza\'nski-Sohnius theorem,
which is not applicable for a free theory.  
Strictly speaking, there is no definite proof of this theorem or of the Coleman-Mandula theorem
for an interacting CFT, as is mentioned in the footnote on p.~13 of 
Weinberg's textbook \cite{WeinbergIII}. This is because the proofs of these theorems are phrased
in terms of the S-matrix, which is ill-defined for CFTs.  
Therefore there is a logical possibility that 
the Coleman-Mandula theorem
and the Haag-\L opusza\'nski-Sohnius theorem, 
instead of  the $a$-theorem, is violated at the AD points under consideration.

In this respect, we think it worthwhile to stress that in a string/M-theory setup 
or in dimensional deconstruction it is quite common to have a new spacetime direction
generated at a particular point in the moduli space, 
thus `violating' the Coleman-Mandula theorem because of the appearance of
a new symmetry which does not commute with the original spacetime symmetry.
The failure of the theorem occurs exactly as anticipated by Coleman-Mandula \cite{ColemanMandula} --
by the appearance of an infinite number of light states, which are the Kaluza-Klein towers
from the point of view of the lower-dimensional theory.
}

A rather trivial question the reader might have is about the decoupled sector in the infrared.  At the AD points with $N_f < 2N_c$, the number $r$ of vector multiplets which couple to mutually nonlocal states is smaller than the rank $N-1$ of the original gauge theory, and so there are $N-1-r$ decoupled free vector multiplets.  
In the calculation in Sec.~\ref{calculation} we actually did not include the contribution of these decoupled vector multiplets to $a$ and $c$. Could their inclusion change the value of $a$ sufficiently so that the $a$-theorem is saved? 
The answer is no --- including them
makes the violation of the $a$-theorem worse, not better. Furthermore,
the violation we found is of order $O(N^2)$
and there are at most $O(N)$ free decoupled vector multiplets, so they cannot change the big picture.

Before closing the paper we would like to briefly reiterate how the established cases of the $a$-theorem fail to apply to our counterexample.
Proofs based on holography are not applicable here,
because our AD points do not have holographic duals which are weakly curved. 
Indeed, we found $a$ and $c$ both to be 
of order $N^2$,  but $a/c\to 7/8$ in the large $N$ limit, whereas
 $a/c\to 1$ in the large $N$ limit of any gauge theory with a weakly-curved AdS$_5$ dual.  Another class of proofs, based on a combination of $a$-maximization and 't Hooft anomaly matching, do not apply, because there are no other symmetries with which U$(1)_R$ can mix, as is required for such proofs to work. 

We hope that these remarks remove any doubts that we have indeed found a counterexample to the $a$-`theorem'. Our finding highlights the peculiar dynamics of the AD points
found in \cite{EHIY} when the ratio $N_f/N_c$ is large. We would deem further study of these SCFTs and the flows between them worthwhile.
The counterexample of lowest rank is the flow from the AD point of the SU$(5)$ theory with eight quarks, to the SU$(4)$ theory with eight massless quarks close to the infinite coupling point. Now, the SU$(3)$ theory with six massless quarks is known to be dual to an SU$(2)$ gauge theory coupled to an $N_f=1$ fundamental hypermultiplet and to the exceptional rank-1 SCFT with flavor symmetry $E_6$ \cite{AS}.  Extending this duality to the SU$(4)$ theory with eight quarks might shed new light on the dynamics of the flow between these two superconformal points.

\acknowledgments
The authors would like to
thank  P. Argyres, V. Balasubramanian, M. R. Douglas, T. Eguchi,  K. Intriligator, D. Martelli, Y. Nakayama,
N. Seiberg, B. Wecht, and E. Witten for  illuminating discussions.
They would like to thank P. Argyres in particular
for careful reading of the manuscript and for his valuable comments.
They would also like to thank  I.~V.~Melnikov for pointing out errors in \eqref{error1} and \eqref{error2} in the version 1 of this paper.

AS gratefully acknowledges support from the Ambrose Monell Foundation and the Institute for Advanced Study.  The work of AS is also partially
supported by NSF grants PHY-0555444 and PHY-0245214.
The work of YT is in part supported by the Carl and Toby Feinberg fellowship
at the Institute for Advanced Study, and by the United States
DOE Grant DE-FG02-90ER40542.

\appendix

\section{Brief history of the $a$-theorem}
\label{history}
The question of whether a version of the $c$-theorem exists in four dimensions was raised by Cardy \cite{Cardy}, who pointed out that the simplest generalization of Zamolodchikov's $c$-function -- constructed from the two-point function of the stress tensor -- need not be monotonically decreasing along flows in more than two dimensions.  Noting that the 2D $c$-function can be alternatively defined by
\begin{equation}
c\equiv -\frac3\pi \int_{S^2} \vev{T_\mu^\mu}\, \sqrt{g}\, d^2 x
\end{equation}
he proposed defining a $d$-dimensional `$c$-function' proportional to 
\begin{equation}
\int_{S^d} \vev{T_\mu^\mu} \,\sqrt{g}\, d^d x
\end{equation}
In four dimensions, this definition reproduces the function $a$ in \eqref{traceanomaly}, since the Weyl curvature of the $4$-sphere vanishes. Furthermore, it naturally provides a definition of $a$-function
away from the conformal point. 

The $c$-theorem for perturbative fixed points was then proved {\it e.g.}~in \cite{Osborn}.
But the definitive modern approach to the $a$-theorem for SCFTs\footnote{
There have been some analyses without the help of supersymmetry, see
e.g.~\cite{hep-th/9805015}, \cite{hep-th/0103237}. }
was initiated by Anselmi and his collaborators, culminating in the papers \cite{Anselmi1,Anselmi2}.
There it was shown how various central charges are related to coefficients
of operator product expansions of the energy momentum tensors and $R$-currents.
Also uncovered were the relations between 't Hooft anomalies of $R$-currents and the central charges.

Following these works, Intriligator and Wecht \cite{IW} discovered the $a$-maximization procedure,  which  fixes the U$(1)_R$ symmetry as a linear combination of possible U$(1)$ symmetries.
Kutasov {\em et al.}~\cite{Kutasov1}
then showed how operators which apparently hit the unitarity bound can be dealt with
by postulating the appearance of extra accidental U$(1)$ symmetries, leading to a more general proof of the $a$-theorem \cite{Kutasov2,IW2}.
An implicit assumption of their approach is that there should be additional U$(1)$ symmetries, defined along the entire flow, with which U$(1)_R$ can mix.

The AdS/CFT correspondence\cite{Maldacena} offers another approach to the $a$-theorem  \cite{holographic-c,hep-th/9807226}. 
At  leading order in the $1/N$ expansion, 
the central charges $a$ and $c$ are equal, and are 
related to the cosmological constant in the dual AdS$_5$ space  \cite{HenningsonSkenderis}.
The RG flow is related to the flow of the scalars in the 5d space, 
which changes the 5d vacuum energy \cite{GPPZ}. In \cite{holographic-c} it was  shown
that the monotonic decrease in $a$ follows from 
a suitable energy condition in the gravity dual.

One particularly interesting holographic manifestation of the $a$-theorem is the following
\cite{obstruction} (see also Sec.~2.2.3 of \cite{nakayama}).  A natural class of
six-dimensional Calabi-Yau cones is  the set of generalized conifolds \begin{equation}
C_{n}:\qquad x^2+y^2+z^2+w^n=0.
\end{equation} The central charge $a_n$ of the theory on  $N$ D3-branes placed
at the origin of the cone $C_n$ can be found by the methods of \cite{BH};
it satisfies $a_n>a_{n+1}$ when $n$ is sufficiently large.
By considering a deformation of the cone $C_{n+1}$ by $\epsilon w^n$ and
 recalling that the radial direction corresponds to the energy scale,
one concludes that the UV theory is $C_{n+1}$ and the IR is $C_n$. This example 
thus seems to violate the $a$-theorem.  However, although
the generalized conifold $C_n$ with $n\ge 3$ is Calabi-Yau in the sense that
nowhere-vanishing holomorphic 3-form exists,  it does {\em not} admit a Ricci-flat metric, because it violates the so-called Bishop bound \cite{obstruction}.  Thus, a direct contradiction with the $a$-theorem is avoided in this case, because this RG flow does not correspond to a valid supergravity solution. 

Finally let us stress that our counterexample to the $a$-theorem does not mean
the end of the quest for the right $c$-function for 4d CFTs, which measures the number
of degrees of freedom.
Indeed, there is still a good chance that some other quantity,
like the ratio $f(T)/T^4$ of the free energy density to the temperature to the fourth \cite{freeenergy},
might satisfy at least the weak form of the $c$-`theorem.'

\end{document}